# Revealing the high-energy electronic excitations underlying the onset of high-temperature superconductivity in cuprates

Claudio Giannetti[1,*], Federico Cilento[2,3], Stefano Dal Conte[4], Giacomo Coslovich[2,3], Gabriele Ferrini[1], Hajo Molegraaf[5], Markus Raichle[6], Ruixing Liang[6,7], Hiroshi Eisaki[8], Martin Greven[9], Andrea Damascelli[6,7], Dirk van der Marel[10] and Fulvio Parmigiani[2,11]

[1]Department of Physics, Università Cattolica del Sacro Cuore, Brescia I-25121, Italy.
[2]Department of Physics, Università degli Studi di Trieste, Trieste I-34127, Italy.
[3]Laboratorio Nazionale TASC, AREA Science Park, Basovizza Trieste I-34012, Italy.
[4]Department of Physics A. Volta, Università degli Studi di Pavia, Pavia I-27100, Italy.
[5]Faculty of Science & Technology, University of Twente, 7500 AE Enschede, The Netherlands.
[6]Department of Physics & Astronomy, University of British Columbia, Vancouver, British Columbia V6T 1Z1, Canada.
[7]Quantum Matter Institute, University of British Columbia, Vancouver, British Columbia V6T 1Z4, Canada.
[8]Nanoelectronics Research Institute, National Institute of Advanced Industrial Science and Technology, Tsukuba, Ibaraki 305-8568, Japan.
[9]School of Physics and Astronomy, University of Minnesota, Minneapolis, Minnesota 55455, USA.
[10]Département de Physique de la Matière Condensée, Université de Genève, CH1211 Genève, Switzerland.
[11]Sincrotrone Trieste S.C.p.A., Basovizza I-34127, Italy.
*Corresponding author: c.giannetti@dmf.unicatt.it

**In strongly-correlated systems the electronic properties at the Fermi energy ($E_F$) are intertwined with those at high energy scales. One of the pivotal challenges in the field of high-temperature superconductivity (HTSC) is to understand whether and how the high energy scale physics associated with Mott-like excitations ($|E-E_F|>1$ eV) is involved in the condensate formation. Here we show the interplay between the many-body high-energy $CuO_2$ excitations at 1.5 and 2 eV and the onset of HTSC. This is revealed by a novel optical pump supercontinuum-probe technique, which provides access to the dynamics of the dielectric function in $Bi_2Sr_2Ca_{0.92}Y_{0.08}Cu_2O_{8+\delta}$ over an extended energy range, after the photoinduced partial suppression of the superconducting pairing. These results unveil an unconventional mechanism at the base of HTSC both below and above the optimal hole concentration required to attain the maximum critical temperature ($T_c$).**

## INTRODUCTION

The high-$T_c$ copper-oxide (cuprates) superconductors are a particular class of strongly-correlated systems in which the interplay[1] between the Cu-3$d$ and the O-2$p$ states determines both the electronic structure close to the Fermi level as well as the high-energy properties related to the formation of Zhang-Rice singlets[2] (hole shared among the four oxygen sites surrounding a Cu and antiferromagnetically coupled to the Cu spin; energy $E_{ZR}\sim 0.5$ eV) and to the charge-transfer processes[3] ($\Delta_{CT}\sim 2$ eV). One of the unsolved issues of high-$T_c$ superconductivity is whether and how the electronic many-body excitations at high-energy scales are involved in the condensate formation in the under- and overdoped regions of the superconducting dome. Solving this problem would provide a benchmark for unconventional models of high-$T_c$ superconductivity, at variance with the BCS theory of conventional superconductivity, where the opening of the superconducting gap induces a significant rearrangement of the quasiparticle excitation spectrum over an energy range of few times $2\Delta_{SC}$ [Ref. 4].

Since the knowledge of the dielectric function $\varepsilon(\omega)$ provides direct information about the underlying electronic structure, optical spectroscopic techniques have been widely used to investigate high-$T_c$ superconductors[5-17]. Recently, equilibrium techniques revealed a superconductivity-induced modification of the $CuO_2$-planes optical properties involving energy scales in excess of 1 eV (Refs. 7,11,12,14-17). These results suggested a possible superconductivity-induced gain in the in-plane kinetic-energy on the underdoped side of the phase diagram[18-20]. However, the identification of the high-energy electronic excitations involved in the onset of HTSC has remained elusive since they overlap in energy with the temperature-dependent narrowing of the Drude-like peak.

Here we solve this question adopting a non-equilibrium approach to disentangle the ultrafast modifications of the high-energy spectral weight ($SW=_0\int^\infty \sigma_1(\omega)d\omega$, $\sigma_1(\omega)=\varepsilon_2(\omega)\cdot\omega/4\pi i$ being the real part of the optical conductivity) from the slower broadening of the Drude-like peak induced by the complete electron-boson thermalization. Since the superconducting gap ($2\Delta_{SC}$) value is related to the total number of excitations[21], an impulsive suppression[22] of $2\Delta_{SC}$ is achieved by photoinjecting excess quasiparticles in the superconducting $Bi_2Sr_2Ca_{0.92}Y_{0.08}Cu_2O_{8+\delta}$ crystals (Y-Bi2212), through an ultrashort light pulse (pump). Exploiting the supercontinuum spectrum produced by a non-linear photonic crystal fibre[23] we are able to probe the high-energy (1.2-2.2 eV) modifications of the *ab*-plane optical properties at low pump fluences (<10 µJ/cm$^2$), avoiding the complete vaporization of the superconducting phase[22,24]. This novel time-resolved optical spectroscopy (TROS) [see Methods] allows us to demonstrate the interplay between many-body electronic excitations at 1.5 and 2 eV and the onset of superconductivity both below and above the optimal hole concentration required to attain the maximum $T_c$.

**RESULTS**

**Equilibrium optical properties of $Bi_2Sr_2Ca_{0.92}Y_{0.08}Cu_2O_{8+\delta}$**

The interpretation of TROS relies on the detailed knowledge of the equilibrium $\varepsilon(\omega)$ of the system. In Fig. 1a, we report the real ($\varepsilon_1$) and imaginary ($\varepsilon_2$) parts of the dielectric function, measured on an optimally-doped sample ($T_c$=96 K, labeled OP96) by spectroscopic ellipsometry at 20 K [Methods]. Below 1.25 eV, $\varepsilon(\omega)$ is dominated by the Drude response of free carriers coupled to a broad spectrum of bosons $I^2\chi(\Omega)$ (Refs. 25,26). In the high-energy region ($\hbar\omega$>1.25 eV), the interband transitions dominate. The best fit to the data (solid black lines) is obtained by modelling the equilibrium dielectric function as a sum of an extended-Drude term ($\varepsilon_D$) and Lorentz oscillators at 0.5 eV and higher energies ($\varepsilon_L$, indexed by i): $\varepsilon_{eq}(T,\omega)=\varepsilon_D(T,\omega)+\sum_i\varepsilon_{L_i}(T,\omega)$ [see Supplementary Note 1 for the details]. The high-energy region of $\varepsilon(\omega)$ is characterized by interband transitions at $\omega_{0_i}$ ~1.46, 2, 2.72 and 3.85 eV. The attribution of the interband transitions in cuprates is a subject of intense debate. As a general phenomenological trend, the charge-transfer (CT) gap edge (hole from the upper Hubbard band with $d_{x^2-y^2}$ symmetry to the O-$2p_{x,y}$ orbitals, see Fig. 1b,c) in the undoped compounds is about 2 eV (Ref. 5). Upon doping, a structure reminiscent of the CT gap moves to higher energies, while the gap is filled with states at the expense of spectral weight at $\hbar\omega$>2 eV (Ref. 5, see Fig. 1d). Dynamical mean field calculations of the electron spectral function and of the *ab*-plane optical conductivity for the hole-doped three-band Hubbard model recently found that the

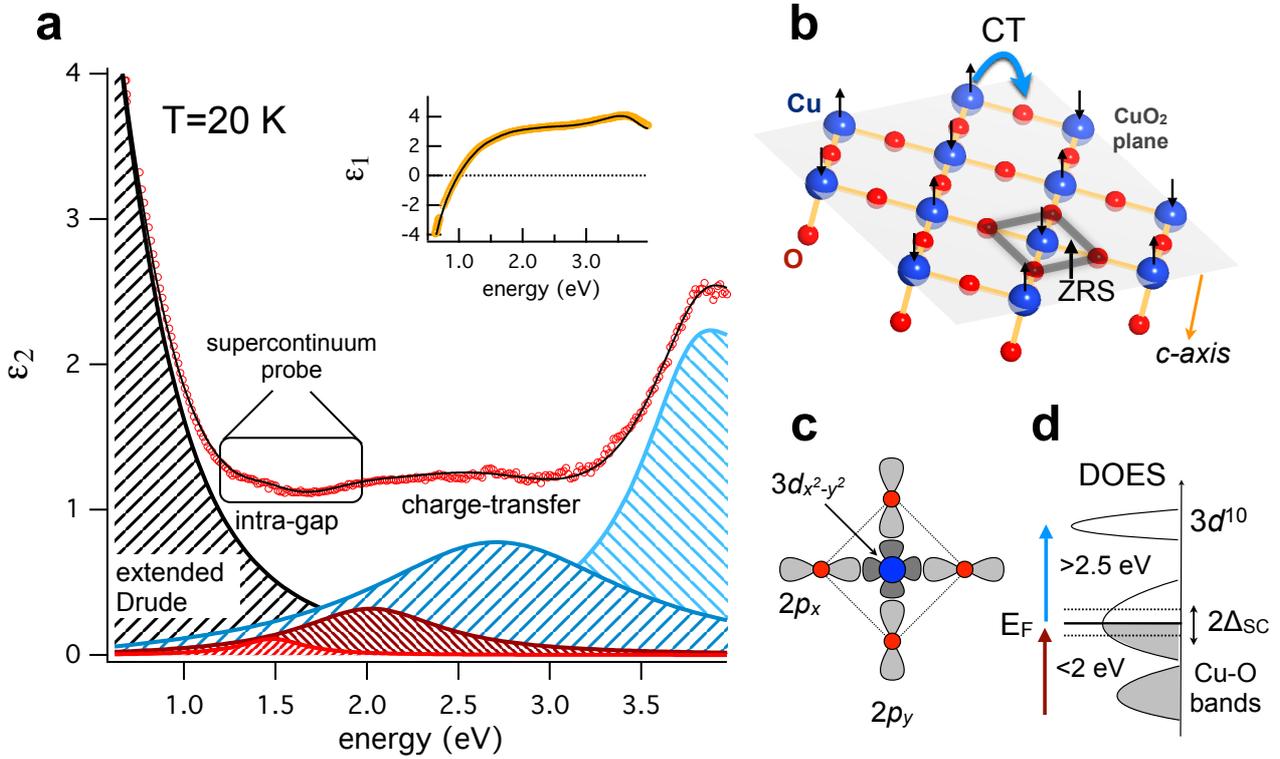

**Figure 1. Dielectric function of optimally-doped $Bi_2Sr_2Ca_{0.92}Y_{0.08}Cu_2O_{8+\delta}$. a**, The high-energy region of the imaginary part of the *ab*-plane dielectric function, measured at *T*=20 K for the optimally-doped sample (estimated hole concentration *p=0*.16, Ref. 45), is shown. The thin black line is the fit to the data described in the text. The values of the parameters resulting from the fit are reported in the Supplementary Table S3. The contributions of the individual interband oscillators at $\omega_{0i}$~1.46, 2, 2.72 and 3.85 eV are indicated as colour-patterned areas. The real part of the dielectric function, with the respective fit, is shown in the inset. Supplementary Figure S1 shows the fits to the reflectivity for *T*=300 K and 100 K and the extracted $I^2\chi(\Omega)$. **b**, Schematics of the Zhang-Rice singlet (ZRS) and of the in-plane charge-transfer (CT) process between Cu and O atoms. **c**, The atomic orbitals involved in the charge-transfer and ZRS are schematically indicated. **d**, Cartoon of the electron density of states (DOES) (Ref. 27). The red and blue arrows represent transitions between mixed Cu-O states and involving a $3d^{10}$ configuration in the final state, respectively.

Fermi level moves into a broad (~2 eV) and structured band of mixed Cu-O character, corresponding to the Zhang-Rice singlet states[27]. The empty upper Hubbard band, which involves Cu-$3d^{10}$ states, is shifted to higher energies with respect to the undoped compound, accounting for the blue-shift of the optical CT edge to 2.5-3 eV. The structures appearing in the optical conductivity at 1-2 eV, i.e., below the remnant of the CT gap, are mostly related to transitions between many-body Cu-O states at binding energies as high as 2 eV (e.g. singlet states) and states at the Fermi energy.

**Time-resolved optical spectroscopy**

In Figure 2 we report the time- and frequency-resolved reflectivity variation in the 1.2-2 eV spectral range ($\delta R/R(\omega,t)$ =$R_{neq}(\omega,t)$ /$R_{eq}(\omega,t)$-1, where $R_{neq}(\omega,t)$ and $R_{eq}(\omega,t)$ are the non-equilibrium (pumped) and equilibrium (unpumped) reflectivities), for the normal (top row), pseudogap (middle row), and superconducting (bottom row) phases at three different dopings. In the normal state, $\delta R/R(\omega,t)$ is positive, with a fast decay over the whole spectrum, and independent of doping. At *T*=100 K, our frequency-resolved measurement reveals a more structured ω-dependence of $\delta R/R(\omega,t)$. In under- and optimally-doped samples a positive variation is measured below ~1.35 eV, while a negative signal, with a fast decay time (~0.5 ps), extends up to 2 eV. The temperatures at which the high-energy negative signal appears ($T^*$~220 K for UD83, $T^*$~140 K for OP96) linearly decrease as the doping increases and correspond to the onset of the universal pseudogap phase with broken time-reversal symmetry[28]. In the overdoped sample, the negative response is nearly absent, while the positive structure at 1.3 eV persists. Below $T_c$, a slow $\delta R/R(\omega,t)$ dynamics appears. This response is strongly doping-dependent, reversing sign when moving from the under- to the overdoped samples. The time-traces of the $\delta R/R(\omega,t)$ scans at 1.5 eV probe energy (see Fig. 2) exactly reproduce the time-resolved reflectivities obtained in the standard single-colour configuration[29,30], i.e., with fixed probe wavelength. The sudden increase of the decay time in the superconducting phase is generally attributed to a bottleneck effect in which the dynamics of quasiparticles is dominated by gap-energy bosons ($\hbar\Omega \sim 2\Delta_{SC}$) emitted during the recombination of Cooper pairs[31-33]. The experimental evidence of this bottleneck in a d-wave superconductor, where the nodal quasiparticles do not emit finite-energy bosons, strongly suggests that the dynamics is dominated by antinodal excitations[31] and is characterized by a transient non-thermal population in the k-space, as recently shown by time-resolved photoemission experiments[34]. These results are a benchmark for the development of realistic non-equilibrium models[6,35,36] of strongly-correlated superconductors, aimed at clarifying the nature of the bosonic glue responsible for high-$T_c$ superconductivity.

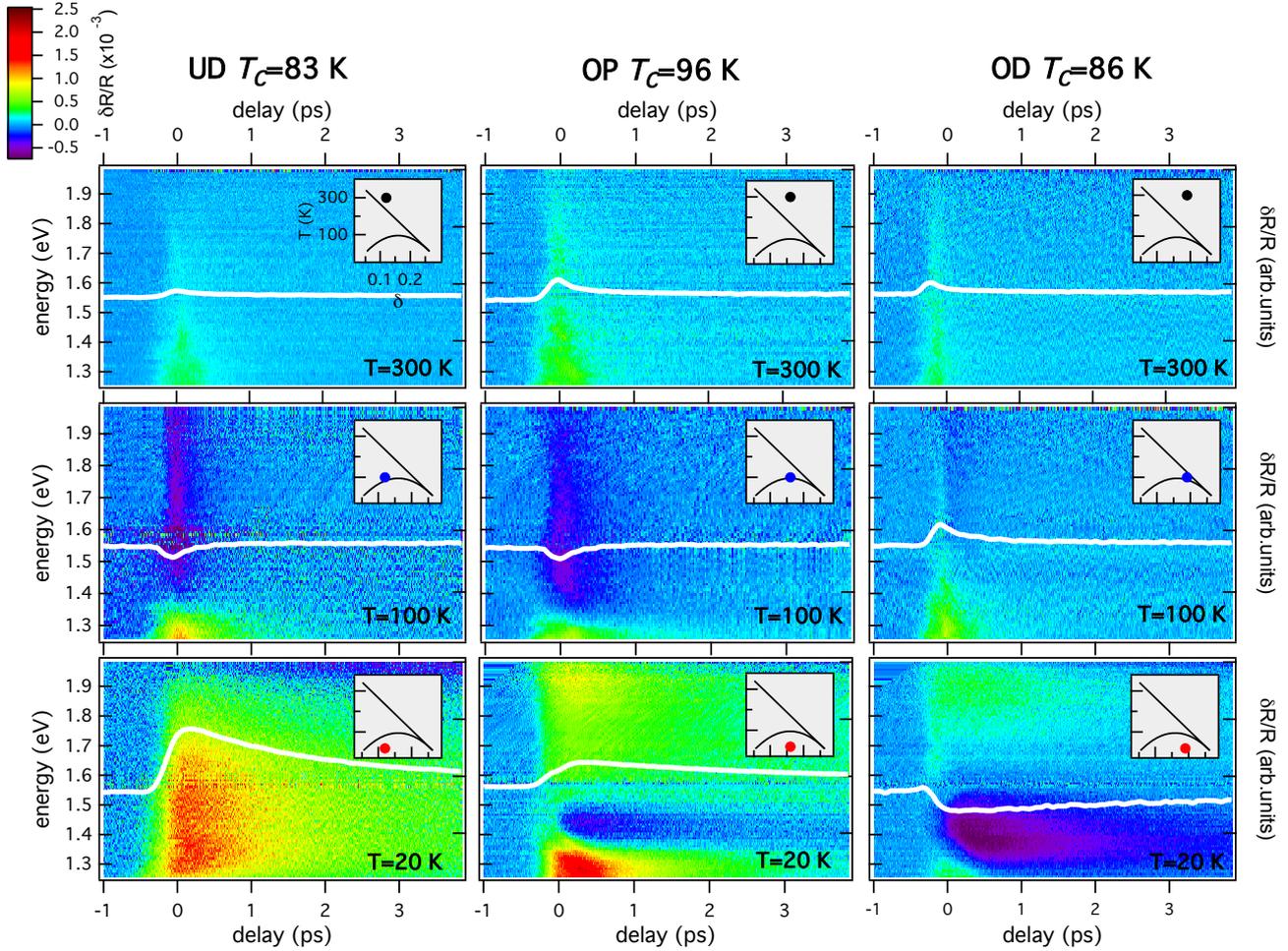

**Figure 2. Energy- and time-resolved reflectivity on $Bi_2Sr_2Ca_{0.92}Y_{0.08}Cu_2O_{8+\delta}$.** The dynamics of the reflectivity is measured over a broad spectral range. The two-dimensional scans of $\delta R/R(\omega,t)$ are reported for three different doping regimes (first column: underdoped (UD), $T_c$=83 K; second column: optimally doped (OP), $T_c$=96 K; third column: overdoped (OD), $T_c$=86 K), in the normal (first row), pseudogap (second row) and superconducting phases (third row). See Methods for estimation of hole content $p$. The insets display schematically the position of each scan in the $T$-$p$ phase diagram of $Bi_2Sr_2Ca_{0.92}Y_{0.08}Cu_2O_{8+\delta}$. The white lines (right axes) are the time-traces at 1.5 eV photon energy.

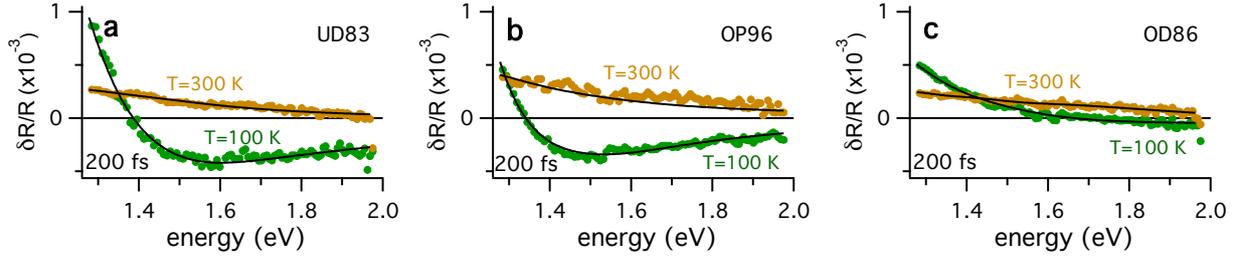

**Figure 3. Energy-resolved spectra and differential fit in the normal and pseudogap phases. a,b,c,** The maximum $\delta R/R(\omega,t)=R_{neq}(\omega,t)/R_{eq}(\omega,t)-1$ ($R_{neq}(\omega,t)$ and $R_{eq}(\omega,t)$ being the non-equilibrium (pumped) and equilibrium (unpumped) reflectivities), measured in the normal (T=300 K, yellow dots) and pseudogap (T=100 K, green dots) phases, is reported for different doping regimes (underdoped (UD83), $T_c$=83 K; optimally doped (OP96), $T_c$=96 K; overdoped (OD86), $T_c$=86 K). The solid lines are the differential fit to the data. The $\delta R/R(\omega,t)$ related to the effective temperature increase is shown in the Supplementary Figure S2. In both the normal and the pseudogap phases the maximum of the fast reflectivity variation is measured at t=200 fs, i.e., directly after all the energy of the 140 fs pump pulses has been delivered to the system.

**Normal and pseudogap phases**

Good spectral resolution is mandatory to obtain a reliable fit of a differential dielectric function $\delta\varepsilon(\omega,t)=\varepsilon_{neq}(\omega,t)-\varepsilon_{eq}(\omega,t)$, where $\varepsilon_{neq}(\omega,t)$ and $\varepsilon_{eq}(\omega,t)$ are the non-equilibrium and equilibrium dielectric functions, to the measured $\delta R/R(\omega,t)$ [see Supplementary Note 2]. In Figure 3a,b,c we report the maximum $\delta R/R(\omega,t)$, measured in the normal and pseudogap phases at t=200 fs. At T=300 K the fit to the data (black lines) is obtained by assuming an increase of the effective temperature, for all the doping levels. As the pseudogap phase is entered, the measured $\delta R/R(\omega,t)$ becomes incompatible with a simple heating of the system. The nearly flat negative signal above ~1.5 eV is well reproduced assuming an impulsive modification of the extended Drude model parameters, such as a weakening of the electron-boson coupling, without invoking any variation of the interband oscillators.

**Superconducting phase**

In Figure 4a we report the maximum $\delta R/R(\omega,t)$, measured in the superconducting phase in the 1.2-2.2 eV energy range. At all doping levels, the data cannot be reproduced by modifying the extended Drude model parameters. The structured variation of the reflectivity at high energies can be only accounted for by assuming a modification of both the 1.5 eV and 2 eV oscillators [see Supplementary Note 2]. The fits to the data automatically satisfy the Kramers-Kronig (KK) relations, since they are obtained as a difference between KK-constrained Lorentz oscillators, and are used to calculate the relative variation of the optical conductivity ($\delta\sigma_1/\sigma_1(\omega,t)$ shown in the inset of Figure 4a). The trend from positive $\delta\sigma_1/\sigma_1(\omega,t)$ in the underdoped to a slightly negative $\delta\sigma_1/\sigma_1(\omega,t)$ in the overdoped samples reveals that the interband spectral weight variation ($\delta SW_{tot}=\delta\omega^2_p(1.5)/8+\delta\omega^2_p(2)/8$, $\omega_p(1.5)$ and $\omega_p(2)$ being the plasma frequencies of the 1.5 and 2 eV oscillators) strongly depends on the doping. We remark that, at higher probe energies (3.14 eV), a negligible $\delta R/R(t)$ value for OP96 is observed, confirming that the transitions at energies larger than 2 eV are not significantly affected by the $2\Delta_{SC}$ suppression.

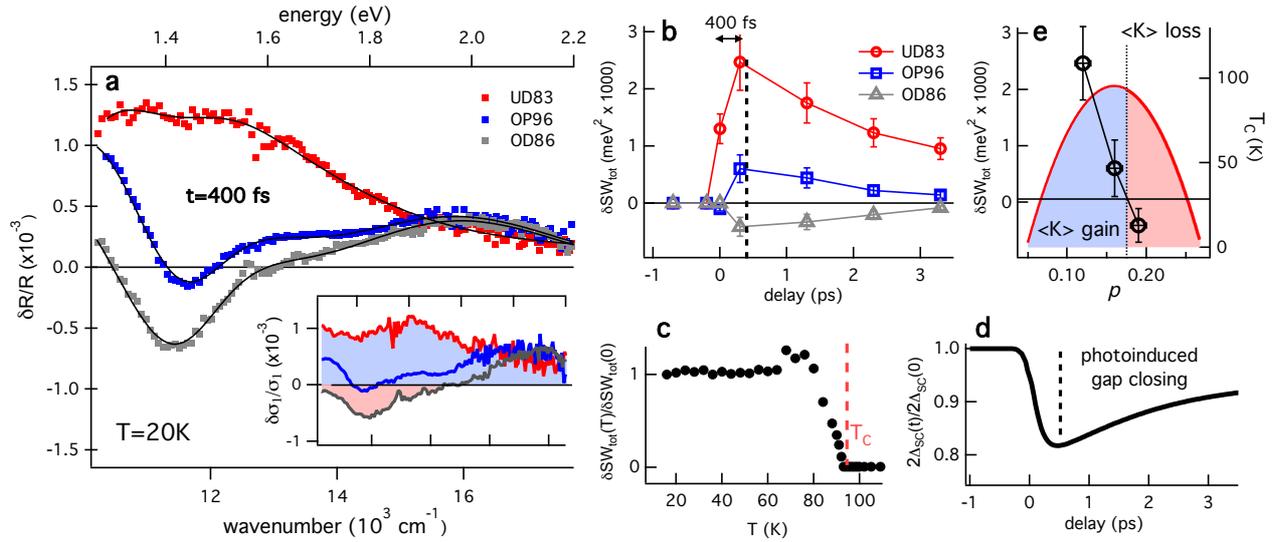

**Figure 4. Energy-resolved spectra and differential fit in the superconducting phase. a,** The $\delta R/R(\omega,t)$ at $t=400$ fs, i.e., the delay at which the maximum signal is measured, is shown for three different dopings in the superconducting phase ($T=20$ K). The black solid lines are differential fits to the data obtained assuming a modification of the 1.5 eV and 2 eV interband transitions. The values of the fitting parameters for the optimally doped sample (OP96) are reported in the Supplementary Table S4. The inset displays the relative variation of the optical conductivity for the three dopings, obtained from the data. The scale of the horizontal axis is the same than the main panel. **b,** The spectral weight variation, $\delta SW_{tot}=\delta SW_{1.5eV}+\delta SW_{2eV}=\delta\omega^2_p(1.5)/8+\delta\omega^2_p(2)/8$ ($\omega_p(1.5)$ and $\omega_p(2)$ being the plasma frequencies of the interband Lorentz oscillators at 1.5 and 2 eV), is reported at different delays for the three dopings. The maximum value of $\delta SW_{tot}$ corresponds to the minimum $\Delta_{SC}$ value at ~400 fs, i.e., after a partial electron-boson thermalization. The error bars represent the standard deviation obtained from the fit. **c,** The $\delta SW_{tot}$ value, relative to the extrapolated zero-temperature value, is estimated from single-colour measurements and reported as a function of the temperature for OP96. Similar results are obtained for UD83 and OD86. **d,** The dynamics of the superconducting gap, assuming the proportionality between $\delta R/R(\omega,t)$ and the photoexcited quasiparticle density, is reported[21,22] (see Supplementary Note 4). At a pump fluence of 10 μJ/cm$^2$, the maximum gap decrease is ~20% at 400 fs delay time. **e,** The black circles represent the maximum $\delta SW_{tot}=\delta SW_{1.5eV}+\delta SW_{2eV}$, observed at 400 fs, as a function of the doping level. The error bars take into account the stability of the differential fit on the equilibrium dielectric function [see Supplementary Notes 1,2]. The left axis has the same units as panel b. The $T_c$-$p$ phase diagram is reported on the right-top axes.

**DISCUSSION**

In the simple energy-gap model for conventional superconductors[37,38], small changes of the interband transitions, over a narrow frequency range of the order of $\omega_{0i}\pm\Delta_{SC}/\hbar$, can arise from the opening of the superconducting gap at the Fermi level. In contrast to this model, the partial suppression of $2\Delta_{SC}$ photoinduced by the pump pulse, induces a change of $R(\omega,t)$ over a spectral range (~1 eV) significantly broader than $2\Delta_{SC}$~80 meV (Ref. 39). This result reveals a dramatic superconductivity-induced modification of the Cu-O electronic excitations at 1.5 and 2 eV.

In the time-domain, we obtain that the dynamics of $\delta SW_{tot}$ (reported in Figure 4b) is nearly exponential with a time constant τ=2.5±0.5 ps. The $\delta SW_{tot}$ variation is completely washed out at longer times (>5 ps), when the complete electron-boson thermalization broadens the Drude peak, overwhelming the contribution of $\delta SW_{tot}$. It is worth noting that the photoinduced modifications of $\delta SW_{tot}$ vanish as $T_c$ is approached from below (Figure 4c), demonstrating that this effect is exclusively related to the impulsive partial suppression of $2\Delta_{SC}$. Further evidence of the direct relation between $2\Delta_{SC}$ and $\delta SW_{tot}$ is obtained comparing our results to the outcomes of time-resolved experiments with probe energy in the mid-infrared[40] and THz regions[41,42], directly showing a recovery time of the superconducting gap and condensate ranging from 2 to 8 ps, for different families of cuprates and different experimental conditions (i.e. temperature and pump fluence). The correspondence between these timescales and the recovery time of $\delta SW_{tot}$ finally demonstrates the interplay between the excitations at 1.5 and 2 eV and the superconducting gap $2\Delta_{SC}$. In addition, our results clarify the origin of the doping- and temperature-dependent $\delta R/R(t)$ signal measured in single-colour experiments[24,29-33] at high-energy scales and demonstrate that the dynamics of $2\Delta_{SC}(t)$ can be reconstructed exploiting the $\delta R/R(t)$ signal measured at high-energies (see Figure 4d and Supplementary Note 4).

Assuming that the total spectral weight change of the interband transitions is compensated by an opposite change of the Drude-like optical conductivity, we can estimate a superconductivity-induced kinetic energy decrease of ~1-2 meV/Cu atom for the underdoped sample [Methods]. This value is very close to the superconductivity-induced kinetic energy gain predicted by several unconventional models[43,44]. As we move from the under- to the overdoped side of the superconducting dome (Fig. 4e), we obtain that the modification of the SW of optical transitions involving excitations at 1.5 and 2 eV, completely accounts for the transition from a superconductivity-induced gain to a loss of kinetic energy[12,17-20], estimated by equilibrium optical spectroscopies, directly measuring the low-energy optical properties, and single-colour time-resolved reflectivity measurements[29]. This result indicates that most of the superconductivity-induced modifications of the excitation spectrum and optical properties at the low-energy scale are compensated by a variation of the in-plane electronic excitations at 1.5 and 2 eV, demonstrating that these excitations are at the base of an unconventional superconductive mechanism both in the under- and overdoped sides of the superconducting dome.

In conclusion, we have exploited the time-resolution of a novel pump supercontinuum-probe optical spectroscopy to unambiguously demonstrate that, in $Bi_2Sr_2Ca_{0.92}Y_{0.08}Cu_2O_{8+\delta}$,

i) the superconductive transition is strongly unconventional both in the under- and overdoped sides of the superconducting dome, involving in-plane Cu-O excitations at 1.5 and 2 eV;

ii) when moving from below to above the optimal hole doping, the spectral weight variation of these features entirely accounts for a crossover from a superconductivity-induced gain to a BCS-like loss of the carrier kinetic energy[4,19,20].

Superconductivity-induced changes of the optical properties at high energy scales ($\hbar\omega > 100\Delta_{SC}$) seem to be a universal feature of high-temperature superconductors[5,7,11], indicating that the comprehension of the interplay between low-energy excitations and the high energy scale physics associated with Mott-like excitations, will be decisive in understanding high-temperature superconductivity.

## METHODS

### Pump supercontinuum-probe technique

The optical pump-probe setup is based on a cavity-dumped Ti:sapphire oscillator. The output is a train of 800 nm-140 fs pulses with an energy of 40 nJ/pulse. A 12 cm long photonic crystal fibre, with a 1.6 µm core, is seeded with 5 nJ/pulse focused into the core by an aspherical lens. The supercontinuum probe output is collimated and refocused on the sample through achromatic doublets in the near-IR/visible range. The 800 nm oscillator output (pump) and the supercontinuum beam (probe), orthogonally polarized, are noncollinearly focused onto the sample. The superposition on the sample and the spot dimensions (40 µm for the IR-pump and 20 µm for the supercontinuum-probe) are monitored through a charge-coupled device camera. The pump fluence is ∼10 µJ/cm$^2$. The probe beam impinges on the sample surface nearly perpendicularly, ensuring that the *ab*-plane reflectivity is measured. The reflected probe beam is spectrally dispersed through a SF11 equilateral prism and imaged on a 128 pixel linear photodiode array (PDA), capturing the 620–1000 nm spectral region. A spectral slice, whose width ranges from 2 nm at 620 nm to 6 nm at 1000 nm, is acquired by each pixel of the array. The probe beam is sampled before the interaction with the pump and used as a reference for the supercontinuum intensity. The outputs of the two PDAs are acquired through a 22 bit/2 MHz fast digitizer, and are divided pixel by pixel to compensate the supercontinuum intensity fluctuations, obtaining a signal to noise ratio of the order of $10^{-4}$ acquiring 2000 spectra in about 1 s. The differential reflectivity signal is obtained modulating the pump beam with a mechanical chopper and performing the difference between unpumped and pumped spectra. The spectral and temporal structure of the supercontinuum pulse has been determined either by exploiting the switching character of the photo-induced insulator-to-metal phase transition in a VO$_2$ multifilm[23] or by standard XFROG techniques. The data reported in Figure 2 have been corrected taking into account the spectral and temporal structure of the supercontinuum pulse.

### Spectroscopic ellipsometry

The *ab*-plane dielectric function of the Bi$_2$Sr$_2$Ca$_{0.92}$Y$_{0.08}$Cu$_2$O$_{8+\delta}$ samples has been obtained by applying the Kramers-Kronig relations to the reflectivity for 50 cm$^{-1}$<ω/2πc<6000 cm$^{-1}$ and directly from ellipsometry for 1500 cm$^{-1}$<ω/2πc<36000 cm$^{-1}$. This combination allows a very accurate determination of ε(ω) in the entire combined frequency range. Owing to the off-normal angle of incidence used with ellipsometry, the *ab*-plane pseudo-dielectric function had to be corrected for the *c*-axis admixture.

### Samples

The Y-substituted Bi2212 single crystals were grown in an image furnace by the traveling-solvent floating-zone technique with a non-zero Y content in order to maximize $T_c$ (Ref. 45). The underdoped samples were annealed at 550 °C for 12 days in a vacuum sealed glass ampoule with copper metal inside. The overdoped samples were annealed in a quartz test tube under pure oxygen flow at 500 °C for 7 days. To avoid damage of the surfaces, the crystals were embedded in Bi$_2$Sr$_2$CaCu$_2$O$_{8+\delta}$ powder during the annealing procedure. In both cases the quartz tube was quenched to ice-water bath after annealing to preserve the oxygen content at annealing temperature. For the optimally-doped sample (OP96) the critical temperature reported ($T_c$=96 K) is the onset-temperature of the superconducting phase transition, the transition being very narrow ($\Delta T_c$<2 K). As a meaningful parameter for the under- ($T_c$=83 K, UD83) and over-doped ($T_c$=86 K, OD86) samples, which have respective transition widths of $\Delta T_c$∼8 K and $\Delta T_c$∼5 K, we report the transition midpoint temperatures. The hole concentration $p$ is estimated through the phenomenological formula[46] $T_c/T_{c,max}$=1−82.6·($p$−0.16)$^2$, where $T_{c,max}$ is the critical temperature of the optimally-doped sample.

### Kinetic energy

Below the critical temperature $T_c$, possible modifications of the high-energy spectral weight $SW_h=\Sigma_i{}_0\int^{\infty}\sigma_{1i}(\omega)d\omega$, including the contribution to the optical conductivity from all the possible interband transitions i, must compensate the spectral weight of the condensate zero-frequency δ-function, $SW_\delta$, and the variation of the spectral weight of the Drude-like optical conductivity, $SW_D = (1/4\pi i)\int_0^\infty \omega \epsilon_D(\omega)d\omega$, related to the free carriers in the conduction band. This is expressed by the Ferrel-Glover-Tinkham (FGT) sum rule[4]:

$$SW_h^N - SW_h^{SC} = SW_\delta - SW_D^N + SW_D^{SC} \tag{1}$$

where the superscripts refer to the normal (N) and superconducting (SC) phases.
In the special case of a single conduction band within the nearest-neighbour tight-binding model[43,44], the total intraband spectral weight $SW_D$ can be related to the kinetic energy $T_\delta$ of the charge carriers (holes) associated to hopping process in the $\delta$ direction, via the relation[43]:

$$\frac{1}{4\pi i}\int_0^\infty \omega \epsilon_D(\omega)d\omega = \frac{\pi^2 a_\delta^2 e^2}{2\hbar^2 V_{Cu}}\langle -T_\delta \rangle \tag{2}$$

where $a_\delta$ is the lattice spacing in the Cu-O plane, projected along the direction determined by the in-plane polarization $\delta$ of the incident light and $V_{Cu}$ is the volume per Cu atom. We obtain $\langle K \rangle = 2\langle T_\delta \rangle$ from the spectral weight variation of the interband oscillators, through the relation:

$$\langle K \rangle = 2\langle T_\delta \rangle = \frac{4\hbar^2 V_{Cu}}{\pi^2 a_\delta^2 e^2}[SW_h^N - SW_h^{SC}] \tag{3}$$

Considering $V_{Cu} = V_{unit\ cell}/8 \sim 1.1\cdot 10^{-22}$ cm$^3$ and $a_\delta = a_{unit\ cell}/\sqrt{2} \sim 3.9$ Å, we obtain that the kinetic energy can be calculated as $\langle K \rangle = 8\hbar^2[SW_h^N - SW_h^{SC}]\cdot(83$ meV/eV$^2)$, where $8\hbar^2[SW_h^N - SW_h^{SC}]$ is the interband spectral weight variation expressed in eV$^2$. A finite value of $SW_h^N - SW_h^{SC}$ thus implies a superconductivity-induced variation of the kinetic energy. To obtain the total kinetic energy variation related to the condensate formation, we estimate a photo-induced breaking of ~3-7% of the Cooper pairs [see Supplementary Note 3] and we extrapolate the measured $SW_h^N - SW_h^{SC}$ values, reported in Figure 4c,d,e, to the values corresponding to the breaking of 100% of the Cooper pairs.

**References**
1. Meinders, M.B.J., Eskes, H., & Sawatzky, G.A. Spectral-weight transfer: Breakdown of low-energy-scale sum rules in correlated systems. *Phys. Rev. B* **48,** 3916 (1993).
2. Zhang, F.C., & Rice T.M. Effective Hamiltonian for the superconducting Cu oxides. *Phys. Rev. B* **37,** 3759-3761 (1988).
3. Zaanen, J., Sawatzky, G.A., & Allen J.W. Band gaps and electronic structure of transition-metal compounds. *Phys. Rev. Lett.* **55,** 418 (1985).
4. Tinkham, M. *Introduction to Superconductivity*. McGraw-Hill, Inc (1996).
5. Basov, D.N. & Timusk, T. Electrodynamics of high-$T_c$ superconductors. *Rev. Mod. Phys.* **77**, 721–779 (2005).
6. Basov, D.N. et al. T. Electrodynamics of Correlated Electron Materials. Accepted for publication on *Rev. Mod. Phys*.
7. Holcomb, M.J., Collman, J.P., & Little, W.A. Optical Evidence of an Electronic Contribution to the Pairing Interaction in Superconducting $Tl_2Ba_2Ca_2Cu_3O_{10}$. *Phys. Rev. Lett.* **73,** 2360 (1994).
8. Tsvetkov, A.A. et al. Global and local measures of the intrinsic Josephson coupling in Tl 2Ba2 CuO 6 as a test of the interlayer tunnelling mode. *Nature* **395**, 360 (1998).
9. Basov, D.N. et al. Sum Rules and Interlayer Conductivity of High-$T_c$ Cuprates. *Science* **283**, 49 (1999).
10. Munzar, D., Bernhard, C., Holden, T., Golnik, A., Humlíček, J., and Cardona, M. Correlation between the Josephson coupling energy and the condensation energy in bilayer cuprate superconductors. *Phys. Rev. B* **64,** 024523 (2001).
11. Rübhausen, M., Gozar, A., Klein, M.V., Guptasarma, P., & Hinks, D.G.. Superconductivity-induced optical changes for energies of 100Δ in the cuprates. *Phys. Rev. B* **63,** 224514 (2001).

**Author contributions**

C.G. and F.P. conceived the project and the time-resolved experiments. C.G. coordinated the research activities with inputs from all the coauthors, particularly F.P., A.D., D.vdM and M.G. C.G., F.C. and G.F. designed and developed the pump supercontinuum-probe technique. C.G., F.C., S. DC and G.C. performed the time-resolved measurements and analysed the data. M.R., R.L., H.E., M.G. and A.D. produced and characterised the crystals. H.M. and D.vdM performed the equilibrium spectroscopic ellipsometry measurements. The text was drafted by C.G. with inputs from F.P., A.D. and D.vdM. All authors extensively discussed the results, the interpretation and revised the manuscript.

**Acknowledgement**

We acknowledge valuable discussion from M. Capone, E. van Heumen, P. Marchetti, F. Carbone and P. Galinetto. F.C., G.C., and F.P. acknowledge the support of the Italian

Ministry of University and Research under Grant Nos. FIRBRBAP045JF2 and FIRB-RBAP06AWK3. The crystal growth work was supported by DOE under Contracts No. DE-FG03-99ER45773 and No. DE-AC03-76SF00515 and by NSF under Grant No. DMR9985067. The work at UBC was supported by the Killam Program (A.D.), the Alfred P. Sloan Foundation (A.D.), the CRC Program (A.D.), NSERC, CFI, CIFAR Quantum Materials, and BCSI.

**Competing Financial Interests**

The authors declare no competing financial interests. Correspondence and requests for materials should be addressed to C.G. (claudio.giannetti@dmf.unicatt.it).

# Supplementary Figures and Tables

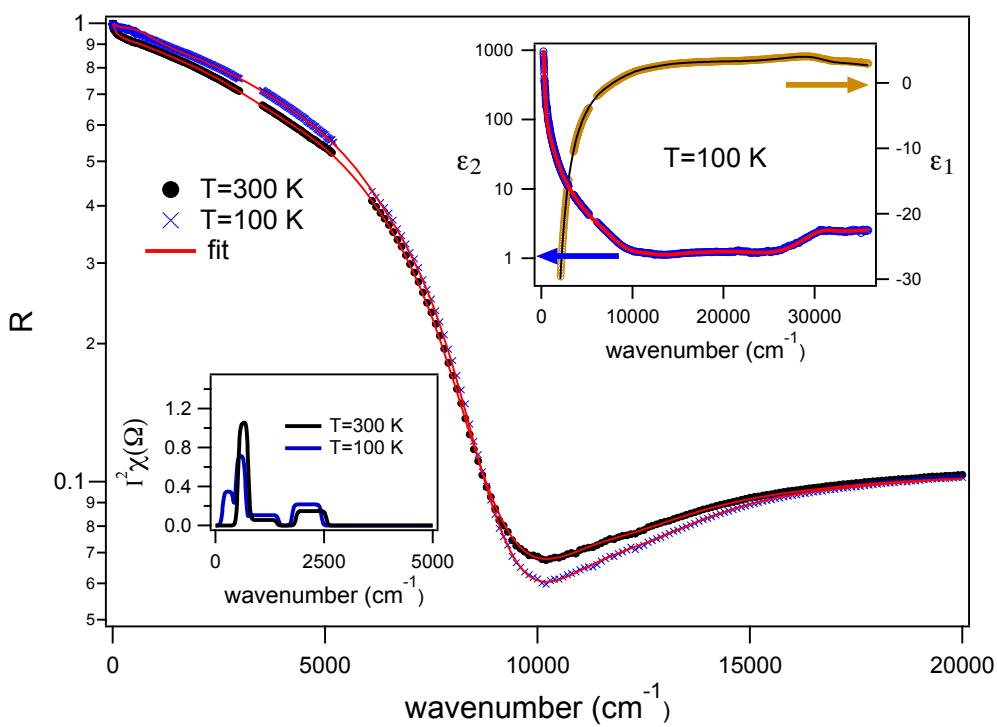

**Supplementary Figure S1. Optical properties of optimally-doped $Bi_2Sr_2Ca_{0.92}Y_{0.08}Cu_2O_{8+\delta}$.**
The reflectivity, measured over a broad frequency range by spectroscopic ellipsometry, is reported for OP96 at $T$=300 K and 100 K. The solid red curves are fits to the data. In the top inset we report $\varepsilon_1$ (right axis) and $\varepsilon_2$ (left axis), measured at $T$=100 K. The solid lines are the fit to the data. In the bottom inset, the boson spectra $I^2\chi(\Omega)$, extracted from the fitting procedure, are reported. All the optical data are reported as a function of the wavenumber ($cm^{-1}$). The conversion to energy units is 8064 $cm^{-1}$=1 eV.

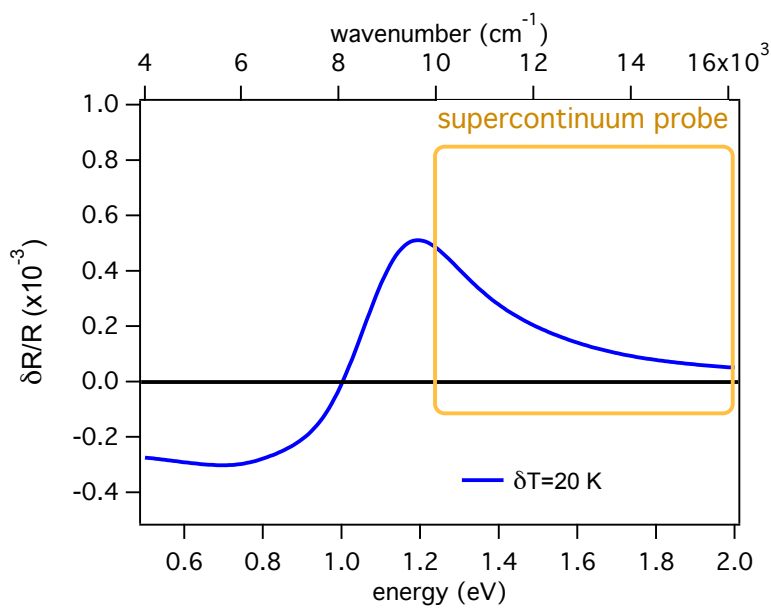

**Supplementary Figure S2. Effective temperature increase.**
The blue solid line is the reflectivity variation calculated for a "quasi-thermal" increase of the temperature $T$, i.e. $\delta T$=20 K.

|  | parameter | T=20 K | T=100 K | T=300 K |
|---|---|---|---|---|
| **extended Drude** | $\varepsilon_{inf}$ | 2.67 | 2.67 | 2.62 |
|  | $\omega_p$ | 17512 | 17650 | 16928 |
|  | $\Gamma_{imp}$ | 53 | 155 | 625 |
| **mid-infrared peak** | $\omega_0$ | 4234 | 4929 | 4264 |
|  | $\omega_{p0}^2$ | 10454991 | 13871222 | 22041800 |
|  | $\gamma_0$ | 3535 | 4706 | 4069 |
|  | $\omega_1$ | 6490 | 6959 | 6789 |
|  | $\omega_{p1}^2$ | 11001998 | 6489852 | 8142014 |
|  | $\gamma_1$ | 3519 | 2949 | 3925 |
| **interband transitions** | $\omega_2$ | 11800 | 11800 | 11650 |
|  | $\omega_{p2}^2$ | 5560610 | 7460610 | 5307060 |
|  | $\gamma_2$ | 3644 | 3944 | 3500 |
|  | $\omega_3$ | 16163 | 16163 | 15409 |
|  | $\omega_{p3}^2$ | 40768500 | 41268500 | 45542000 |
|  | $\gamma_3$ | 8304 | 8304 | 8905 |
|  | $\omega_4$ | 21947 | 21947 | 21300 |
|  | $\omega_{p4}^2$ | 225776000 | 230776025 | 223159025 |
|  | $\gamma_4$ | 13998 | 13898 | 13898 |
|  | $\omega_5$ | 31057 | 31057 | 30756 |
|  | $\omega_{p5}^2$ | 288626121 | 288626121 | 320536896 |
|  | $\gamma_5$ | 6191 | 6191 | 6908 |
|  | $\omega_6$ | 35146 | 35146 | 34946 |
|  | $\omega_{p6}^2$ | 217474009 | 217474009 | 214474009 |
|  | $\gamma_6$ | 6396 | 6396 | 6396 |
|  | $\omega_7$ | 40421 | 40421 | 40421 |
|  | $\omega_{p7}^2$ | 750212100 | 750212100 | 756371984 |
|  | $\gamma_7$ | 7518 | 7518 | 7518 |

**Supplementary Table S3. Parameters of the equilibrium dielectric function.**
Values of the parameters resulting from the fit to $\varepsilon_1(T,\omega)$ and $\varepsilon_2(T,\omega)$ for OP96. All the values are expressed in inverse centimeter units (cm$^{-1}$). The values reported in the table are truncated at the last significative figure, considering the standard deviation given by the fitting procedure.

| | parameters | t=0.4 ps |
|---|---|---|
| **extended Drude** | T | 39 |
| **interband transitions** | $\omega_2$ | 11800 |
| | $\omega_{p2}^2$ | 5679264 |
| | $\gamma_2$ | 3698 |
| | $\omega_2$ | 16176 |
| | $\omega_{p2}^2$ | 40689072 |
| | $\gamma_2$ | 8283 |

**Supplementary Table S4. Parameters of the non-equilibrium dielectric function.**
Values of the parameters modified in the differential dielectric function $\delta\varepsilon(\omega,t) = \varepsilon_{neq}(\omega,t) - \varepsilon_{eq}(\omega,t)$ to fit the data at $T$=20 K for the optimally doped (OD96 sample). The best fit is reported in Fig. 4a of the manuscript.

# Supplementary Note 1

## Equilibrium optical properties of $Bi_2Sr_2Ca_{0.92}Y_{0.08}Cu_2O_{8+\delta}$

In Supplementary Figure S1 we report the reflectivity measured on OP96 by spectroscopic ellipsometry at 300 K and 100 K. The reflectivity measured at 20 K is not reported, since, on the scale of Supplementary Figure S1, it overlaps with the 100 K results. In the top inset to Supplementary Figure S1 we report the real ($\varepsilon_1$) and imaginary ($\varepsilon_2$) parts of the dielectric function, measured on OP96 at 100 K. From these data we can argue that, below 10000 cm$^{-1}$ (1.25 eV), $\varepsilon(\omega)$ is dominated by the Drude response of free carriers coupled to a broad spectrum of bosons [25,26], whereas in the high-energy region ($\hbar\omega$>1.25 eV), a major role is played by the interband transitions. The best fit to the data (solid black lines) is obtained modeling the equilibrium dielectric function as:

$$\epsilon_{\text{eq}}(T,\omega) = \epsilon_D(T,\omega) + \Sigma_i \epsilon_{L_i}(T,\omega) \tag{S1}$$

where $\varepsilon_D(T,\omega)$ and $\varepsilon_{L_i}(T,\omega)$ represent Drude and Lorentz oscillators, indexed by i. The reflectivity is obtained from the dielectric function through the relation:

$$R(T,\omega) = \frac{1 - \sqrt{\epsilon(T,\omega)}}{1 + \sqrt{\epsilon(T,\omega)}} \tag{S2}$$

The Drude component of the dielectric function is given by the extended Drude model [25,26]:

$$\epsilon_D(\omega) = -\frac{\omega_p^2}{\omega(\omega + M(\omega,T))} \tag{S3}$$

where $M(\omega,T)$ is the temperature-dependent memory function, given by:

$$M(\omega,T)) = \Gamma_{imp} - 2i \int_0^\infty d\Omega K(\omega,\Omega;T) I^2\chi(\Omega) \tag{S4}$$

where $\Gamma_{imp}$ is the impurity scattering rate and $I^2\chi(\Omega)$ is the spectrum of the bosons coupled to the electrons through the kernel function $K(\omega,\Omega;T)$. The kernel function is calculated by [47]:

$$K(x,y) = \frac{i}{y} + \left\{ \frac{y-x}{x} \left[ \Psi(1-ix+iy) - \Psi(1+iy) \right] \right\} - \{y \to -y\} \tag{S5}$$

where $x=\omega/2\pi T$ and $y=\Omega/2\pi T$ and $\Psi(x,y)$ is the digamma function. In the bottom inset to Supplementary Figure S1 we report the boson spectra $I^2\chi(\Omega)$ extracted from the fit at $T$=300 K and $T$=100 K. The spectra are characterized by a narrow peak at 50-70 meV and a broad spectrum extending up to 400 meV, in agreement with the literature [25,26]. Below $T_c$ the far-infrared reflectivity is dominated by the opening of the superconducting gap and by the emergence of the condensate $\delta(0)$ function. For this reason, the extraction of the boson spectral function below $T_c$ is difficult and is still subject of debate [26]. To fit the reflectivity at $T$=20 K we used the $I^2\chi(\Omega)$ determined at $T$=100 K. Neglecting a possible sharp decrease below $T_c$ of the scattering time [48] does not alter the conclusions of this work.

To satisfactorily reproduce the optical properties both in the underdoped regime, characterized by strong mid-infrared (MIR) peaks, and in the overdoped regime, where the MIR peaks broadens and can be completely accounted for by the extended Drude model, we used a general dielectric function model where both MIR peaks and the $I^2\chi(\Omega)$ glue are present.

In Supplementary Figure S1, the solid lines are the fit to the data. In Supplementary Table S1 we report all the parameters obtained from the fitting procedure on OP96, at $T$=300 K, $T$=100 K and $T$=20 K.

The best fit to the full spectrum is obtained by including in the fit two oscillators at 0.5 eV and 0.8 eV. The spectral weight of these peaks represents less than 10 percent of the free particle spectral weight and depends strongly on temperature. For this reason this part of the spectrum is part of the high energy range of the generalized Drude peak. Correspondingly, the energy range and intensity of $I^2\chi(\Omega)$ is higher than shown in the inset of Supplementary Figure S1 [26]. The representation of part of the generalized Drude response by two oscillators at 0.5 eV and 0.8 eV was used for convenience of rapid convergence of the fitting routine, and does not affect the conclusions of this work.

The interband transitions in the near-IR/visible/UV spectral range are reproduced using six Lorentz oscillators at $\omega_{0i} \sim$1.46, 2, 2.72, 3.85, 4.4 and 5 eV. The number of the interband oscillators has been fixed to the minimum necessary to obtain a stable fit. Adding more oscillators does not significantly improve the $\chi^2$ of the fit in the 1-5 eV region.

The attribution of the 1.46, 2, 2.72, 3.85 eV transitions, that are the most relevant to this work, is disussed in the manuscript. Interpretation of the 1.46 eV structure in terms of a *d-d* excitation [49] can be ruled out, since the oscillator strength obtained from the fit is incompatible with the temperature dependence of a phonon assisted *d-d* transition, observed in undoped compounds [50].

The dielectric functions of UD83 and OD86 have been extrapolated from the $\varepsilon_{eq}(T,\omega)$ of OP96, following the trend of the optical properties at different dopings, reported in Ref. 20.

# Supplementary Note 2

### Differential model and non-equilibrium dielectric function

In a time-resolved experiment the impulsive photoinjection of excitations is achieved by an ultrashort laser pulse (FWHM=140 fs) in the near-IR spectral region (λ=800 nm). The recovery dynamics of the non-equilibrium (pumped) electronic distribution can be roughly separated in three timescales:

i) **0-50 fs electron thermalization**
The pump pulse creates a non-thermal population, injecting excitations at ~1.5 eV. The scattering rate, obtained through optical measurements, is strongly determined by both electron-electron and electron-phonon scattering processes and is roughly proportional to both the frequency and temperature, i.e., $1/\tau(\omega,T)\propto\omega+T$ [51]. At 1.5 eV energy, the frequency-dependent scattering rate is ~4000 cm$^{-1}$, corresponding to lifetime of ~1 fs. In the first tens of femtoseconds, the non-equilibrium electrons lose energy through a cascade process and low-energy excitations are accumulated at the top of the gap. As the electronic excitations decrease their energy, the scattering time proportionally decreases. The non-equilibrium population created within the 140 fs pulse duration can be assumed as "quasi-thermal", being described by an effective chemical potential ($\mu_{eff}$) and temperature ($T_{eff}$). In this frame, the photoinduced non-equilibrium population created during the pulse duration (140 fs) is independent of the particular pump photon energy. To test this prediction, we repeated the $\delta R/R(\omega,t)$ measurements with a photon pump energy of 3.14 eV, obtaining the same results.

The density of photoinjected excitations can be roughly estimated considering the absorbed laser fluence of 10 µJ/cm$^2$ per pulse, corresponding to an absorbed energy density of 0.6 J/cm$^3$, $\lambda_d$=160 nm being the penetration depth at 800 nm wavelength. Considering a volume per Cu atom of 1.13·10$^{-22}$ cm$^3$ and assuming that the number of gap-energy excitations produced by each 1.5 eV-photon is approximately ~$\hbar\omega/2\Delta_{SC}$=1.5 eV/0.04 eV~40, the density of broken Cooper pairs is ~5.5·10$^{-3}$ per Cu atom. This value is about the 7% of the superfluid density in optimally-doped Bi2212 [52]. This calculation does not take into account the d-wave character of the superconducting gap and the possibility that energy can be released to bosons before the complete electron thermalization. A more precise estimation of the density of photoinjected excitation, based on a phenomenological approach, is reported in Supplementary Note 3.

ii) **50-500 fs electron and boson thermalization**
On this timescale the quasiparticles (QPs) exchange energy with the boson population at the equilibrium temperature $T_{eq}$. The energy exchange can be related either to a direct inelastic scattering process between gap-energy excitations and bosons or to the selective emission of bosons during the recombination of QPs to reform Cooper pairs. On the sub-ps timescale the physical scenario is that of a non-equilibrium population of QPs thermalized with a subset of bosonic modes at a temperature larger than $T_{eq}$.

iii) **>500 fs boson thermalization**
On this timescale the subset of bosons strongly coupled to the QPs thermalizes with the boson thermal reservoir through inelastic boson-boson scattering processes.

This schematic physical picture is confirmed by recent results obtained by different techniques, such as time-resolved photoemission [53], time-resolved electron-diffraction [54] and time-resolved Raman scattering [55].

The reflectivity variations ($\delta R/R(\omega,t)=R_{neq}/R_{eq}(\omega,t)-1$) measured by the TROS technique are related to the variation of the dielectric function $\delta\varepsilon(\omega,t)=\varepsilon_{neq}(\omega,t)-\varepsilon_{eq}(\omega,t)$. Fitting a differential dielectric function to the measured $\delta R/R(\omega,t)$, allows us to obtain the dynamics of $\varepsilon_{neq}(\omega,t)$.

In Supplementary Figure S2 (blue line) we report the calculated differential reflectivity, i.e. $\delta R=R(\omega,T+\delta T)/R(\omega,T)-1$, as the effective temperature of the system is changed of $\delta T$=20 K. Qualitatively a "quasi-thermal" temperature variation induces a positive change of the reflectivity above 1 eV and a negative variation below 1 eV. Above 1.2 eV the reflectivity variation amplitude monotonically decreases as the photon energy increases. This shape of the reflectivity variation is typical of a temperature-related broadening of the Drude peak. While an effective temperature variation is assumed to satisfactorily fit the data at $T$=300 K (see Figure 3), this effect cannot account for the frequency-resolved reflectivity variations measured in both the pseudogap (PG, $T$=100 K) and the superconducting (SC, $T$=20 K) phases. The signals measured in the PG and SC phases are thus the fingerprint of a genuine modification of the dielectric function as excitations are photo-injected, well beyond a simple broadening of the Drude peak. In the PG, the flat negative signal above ~1.4 eV (see Figure 3a,b,c) is well reproduced assuming an impulsive modification of the extended Drude model parameters, such as a weakening of the electron-boson coupling, without invoking any variation of the interband oscillators.

To reproduce the reflectivity variation in the SC phase (see Figure 4a) we must assume a modification of the first two interband oscillator ($\omega_2$ and $\omega_3$ in Supplementary Table S1) at 11800 cm$^{-1}$ (1.46 eV) and 16163 cm$^{-1}$ (2 eV). The conservation of the total spectral weight is guaranteed by the constraint that the sum of the squared plasma frequencies of the extended Drude model and of the interband oscillators is constant. In Supplementary Table S2 we report the parameters modified in $\varepsilon_{neq}(\omega,t)$ to obtain the best fit to the data, reported in Fig. 4a of the manuscript. The results of the differential fitting procedure are very stable on the choice of the equilibrium dielectric function, obtained in the Supplementary Note 1. The same results are obtained assuming a different $\varepsilon_{eq}(\omega,t)$ (for example with a different number of interband oscillators or a different glue function). For this reason, the equilibrium dielectric function used can be considered as a "realistic" dielectric function, even if, possibly, not the best dielectric function one can get (since the procedure to obtain it is often questionable). In Figure 4e, the error bars indicate the range of $\delta SW_{tot}$ values that can be obtained starting form different $\varepsilon_{eq}(\omega,t)$.

The differential dielectric function fitting procedure allows us to conclude that, as $T_c$ is crossed, the $\delta R/R(\omega,t)$ signal measured in the near IR/visible region is associated to a modification of the interband transitions, beyond a simple impulsive broadening of the Drude function. This conclusion holds for all the doping regimes investigated in the present measurements. These results finally shed light on the long-standing question [31,56] about the origin of the $\delta R/R(t)$ measured in the one-colour time-resolved reflectivity experiment [31,32]. The measured signal is not originated by an *excited state absorption*, related to the variation of the electronic distribution within the unvaried electronic bands, but to a real modification of the underlying electronic excitations and, in particular, of the interband transitions at 1.5-2 eV. Nonetheless, the measured $\delta R/R(\omega,t)$ is proportional to the density of photo-injected quasiparticles, as commonly assumed (see Ref. [57] and references therein).

# Supplementary Note 3

## Excitation density

In the Supplementary Note 1 we reported an estimation of the total number of excitations impulsively injected in the system by the pump pulse. However, the calculation does not take into account both the d-wave character of the superconducting gap and the possibility that energy can be released to bosons before the complete electron thermalization [21]. A more precise calculation of the density of photoinjected excitations is based on a phenomenological approach. We take advantage of the upper limit in the photoexcitation density set by the photo-induced phase transition to the normal state reported at high pump fluence on HTSC [57,24,58]. This non-thermal phase transition has a first-order character and takes place at a finite superconducting gap value [22], as predicted by the $\mu_{eff}$ non-equilibrium superconductivity model [59,21,24].
In the Y-Bi2212 we observe a photo-induced phase transition at ~60 μJ/cm$^2$ [22,24]. From numerical calculations, performed by Nicol et al. [21] within the $\mu_{eff}$ model, the instability of the superconducting state is predicted when 15-20% of Cooper pairs are broken.

Extrapolating back to our working fluence (10 µJ/cm$^2$), we can estimate the number of photo-excited Cooper pairs to be around 3%.

## Supplementary Note 4

### Gap suppression

When a superconducting system is strongly perturbed by an ultra-short laser pulse, the superconducting order parameter $\Delta_{SC}$ is expected to exhibit strong variations in time. To obtain the instantaneous gap value during the recovery dynamics we developed a time dependent model based on the Rothwarf-Taylor equations (RTE) [60,22]:

$$\dot{n} = I_{QP}(t) + 2\gamma p - \beta n^2$$
$$\dot{p} = I_{ph}(t) - \gamma p + \beta n^2/2 - \gamma_{esc}(t) \cdot (p - p_T)$$

(S6)

describing the density of excitations $n$ (quasiparticles, QP) coupled to phonons, $p$ being the gap-energy phonon density. The non-equilibrium QPs and phonons are photo-injected in the system through the $I_{QP}(t)$ and $I_{ph}(t)$ terms. A gaussian temporal profile of $I_{QP}(t)$ and $I_{ph}(t)$, with the same time-width as the laser pulse, is assumed. The coupling of the electronic and phonon population is obtained through: a) the annihilation of a Cooper pair via a gap-energy phonon absorption ($\gamma p$ term) and b) the emission of gap-energy phonons during the two-body direct recombination of excitations to form a Cooper pair ($\beta n^2$ term).
In the phonon bottleneck regime ($\gamma > \gamma_{esc}$), the relaxation of the excitations is ultimately regulated by the escape rate of the non-equilibrium gap-energy phonons ($\gamma_{esc}(p-p_T)$ term, $p_T$ being the thermal phonon density). The $\gamma_{esc}$ value is determined both by the escape rate of the non-equilibrium phonons from the probed region and by the energy relaxation through inelastic scattering with the thermal phonons and is directly dependent on the superconducting gap value [22]. The time evolution of $\Delta_{SC}(t)$ can be obtained from the $\delta R/R(t)$ time-traces at 800 nm probe wavelength (see Figure 2), under the following assumptions:
a) The $\delta R/R(t)$ time-trace is assumed to be proportional to the solution $n(t)$ of Eqs. (S6), in agreement with the literature [24,29-33,41,57,61-63].
b) the time-dependent non-equilibrium superconducting gap $\Delta_{SC}(n(t))$ can be expressed as a function of $n(t)$, considering the effective temperature ($T_{eff}$) and chemical potential ($\mu_{eff}$) models [59,21,24,22]. In both cases the normalized $\Delta_{SC}(n(t))$ depends on (1-$a \cdot n(t)^{3/2}$) (being $a$ a conversion factor) for a $d$-wave gap symmetry [21,22].
c) While $\gamma$ and $\beta$ can be assumed as constant during the decay dynamics, $\gamma_{esc}(\Delta_{SC}(n(t)))$ is the only time-dependent parameter.
d) $\gamma_{esc}(t)$ quadratically depends on the instantaneous gap value, as demonstrated in Ref. [22], i.e. $\gamma_{esc}(t) = \gamma_{esc}(0) \cdot [\Delta_{SC}(n(t))/\Delta_{SC}(0)]^2$, $\gamma_{esc}(0)$ being the $\gamma_{esc}$ value corresponding to the unperturbed gap $\Delta_{SC}(0)$.

Fitting the $\delta R/R(t)$ time-traces (Figure 2), measured at 10 µJ/cm$^2$ pump fluence, with the solution of Eqs. (S6) we are able to extract the instantaneous $\Delta_{SC}(t)$ value, reported in Figure 4b. The minimum gap value is obtained at t=400 fs delay and corresponds to ~80% of $\Delta_{SC}(0)$.

# Supplementary References